\documentclass[10pt]{iopart}

\usepackage{graphicx}
\usepackage{dcolumn}
\usepackage{bm}
\usepackage{color}

\begin{document}

\title[ECCD effect]{Modeling of the ECCD injection effect on the Heliotron J and LHD plasma stability}


\author{J. Varela}
\ead{jacobo.varela@nifs.ac.jp}
\address{National Institute for Fusion Science, National Institute of Natural Science, Toki, 509-5292, Japan}
\address{Universidad Carlos III de Madrid, 28911 Leganes, Madrid, Spain}
\author{K. Nagasaki}
\address{Institute of Advanced Energy, Kyoto University, Uji, Japan}
\author{K. Nagaoka}
\address{National Institute for Fusion Science, National Institute of Natural Science, Toki, 509-5292, Japan}
\address{Graduate School of Science, Nagoya University, Nagoya, 464-8602, Japan}
\author{S. Yamamoto}
\address{National Institutes for Quantum and Radiological Science and Technology, Naka, Ibaraki 311-0193, Japan}
\author{K. Y. Watanabe}
\address{National Institute for Fusion Science, National Institute of Natural Science, Toki, 509-5292, Japan}
\author{D. A. Spong}
\address{Oak Ridge National Laboratory, Oak Ridge, Tennessee 37831-8071, USA}
\author{L. Garcia}
\address{Universidad Carlos III de Madrid, 28911 Leganes, Madrid, Spain}
\author{A. Cappa}
\address{Laboratorio Nacional de Fusion CIEMAT, Madrid, Spain}
\author{A. Azegami}
\address{Graduate School of Science, Nagoya University, Nagoya, 464-8602, Japan}

\date{\today}

\begin{abstract}
The aim of the study is to analyze the stability of the Energetic Particle Modes (EPM) and Alfven Eigenmodes (AE) in Helitron J and LHD plasma if the electron cyclotron current drive (ECCD) is applied. The analysis is performed using the code FAR3d that solves the reduced MHD equations describing the linear evolution of the poloidal flux and the toroidal component of the vorticity in a full 3D system, coupled with equations of density and parallel velocity moments for the energetic particle (EP) species, including the effect of the acoustic modes. The Landau damping and resonant destabilization effects are added via the closure relation. The simulation results show that the $n=1$ EPM and $n=2$ Global AE (GAE) in Heliotron J plasma can be stabilized if the magnetic shear is enhanced at the plasma periphery by an increase (co-ECCD injection) or decrease (ctr-ECCD injection) of the rotational transform at the magnetic axis ($\rlap{-} \iota_{0}$). In the ctr-ECCD simulations, the EPM/AE growth rate decreases only below a given $\rlap{-} \iota_{0}$, similar to the ECCD intensity threshold observed in the experiments. In addition, ctr-ECCD simulations show an enhancement of the continuum damping. The simulations of the LHD discharges with ctr-ECCD injection indicate the stabilization of the $n=1$ EPM, $n=2$ Toroidal AE (TAE) and $n=3$ TAE, caused by an enhancement of the continuum damping in the inner plasma leading to a higher EP $\beta$ threshold with respect to the co- and no-ECCD simulations.
\end{abstract}

%
%
%
%
%


\vspace{2pc}

\maketitle

\ioptwocol

\section{Introduction \label{sec:introduction}}

The external injection of electron cyclotron waves (ECW) \cite{1,2} modifies the plasma stability of nuclear fusion devices by the generation of non inductive currents in the plasma. In particular, the electron cyclotron current drive (ECCD) \cite{3,4,5,6,7} can improve the stability of the pressure/current gradient driven modes and the Alfven Eigenmodes (AE) in Tokamaks \cite{8,9,10,11,12,13} and Stellarators \cite{14,15,16,17,18}.

ECCD injection in LHD \cite{19} attains the stabilization of the energetic-ion-driven resistive interchange mode (EIC) \cite{20,21}, Toroidal and global Alfven eigenmodes (TAE / GAE) \cite{22} as well as pressure gradient driven modes \cite{23}. The same way, ECCD injection in Heliotron J \cite{24} leads to the stabilization of AEs \cite{25,26}.

Energetic particle (EP) driven instabilities enhance the transport of fusion produced alpha particles, energetic hydrogen neutral beams and ion cyclotron resonance heated particles (ICRF) \cite{27,28,29}, leading to a lower heating efficiency in nuclear fusion devices \cite{30,31,32}. Alfvenic instabilities are triggered if there is a resonance between the unstable mode frequency and the EP drift, bounce or transit frequencies, enhancing the EP losses.

Heliotron J and LHD are helical devices heated by NBI lines. A tangential NBI is deposited on-axis with an energy of $34$ keV and an injection power of $0.7$ MW in the case of the Heliotron J \cite{33}. Three NBI lines parallel to the magnetic axis are injected in LHD plasma with an energy of $180$ keV and a power of $16$ MW \cite{34}. In addition, the LHD device has two NBIs perpendicular to the magnetic axis injected in the plasma periphery with an energy of 32 keV and a power of $12$ MW \cite{35}.

The aim of the present study is to simulate the stabilization of the energetic particle modes (EPM), Global Alfven Eigenmodes (GAE) and Toroidal Alfven Eigenmodes (TAE) observed in Heliotron J and LHD discharges if there is an ECCD injection \cite{22,25,26}. The simulations are performed using the FAR3D code \cite{36,37,38} that solves the reduced linear resistive MHD equations and the moment equations of the energetic ion density and parallel velocity \cite{39,40}. The numerical model includes the linear wave-particle resonance effects required for Landau damping/growth and the parallel momentum response of the thermal plasma required for coupling to the geodesic acoustic waves \cite{41}. The code variables evolve starting from an equilibria calculated by the VMEC code \cite{42}. The effect of the ECCD is included in the simulation as a modification of the VMEC rotational transform profile.

This paper is organized as follows. The numerical scheme is introduced and the equilibrium properties are described in section \ref{sec:model}. The modeling of the EPM and GAE stabilization in Heliotron J is performed in section \ref{sec:HJ}. Next, the modeling of the EPM and TAE stabilization in LHD is studied in section \ref{sec:LHD}. Finally, the conclusions of this paper are presented in section \ref{sec:conclusions}.

\section{Numerical scheme \label{sec:model}}

The FAR3d codes solves the reduced MHD equations describing the linear evolution of the poloidal flux, the total pressure and the toroidal component of the vorticity, as well the thermal parallel velocity (required to add the effect of the acoustic modes in the simulations) \cite{43}. The thermal plasma equations are coupled with the equations of the EP density and parallel velocity moments. FAR3d applies the stellarator expansion and the main assumptions for the derivation of the set of reduced equations are high aspect ratio, medium $\beta$ (of the order of the inverse aspect ratio), small variation of the fields and small resistivity. The code uses the equilibrium flux coordinates ($\rho$, $\theta$, $\zeta$), finite differences in the radial direction and Fourier expansions in the two angular variables. There are two numerical schemes to resolve the linear equations: a semi-implicit initial value or an eigenvalue solver. The initial value solver calculates the mode with the largest growth rate (dominant mode) and the eigen-solver the stable and unstable modes (dominant + sub-dominant modes). The present model was already used to study the activity of TAEs \cite{43,44} and EIC \cite{20,45,46} as well as the effect of the NBI current drive on the AE stability \cite{47} in LHD plasma, indicating a reasonable agreement with the observations.

\subsection{Equilibrium properties}

Fixed boundary results from the VMEC equilibrium code \cite{42} are calculated for the LHD discharge $138675$ and the Heliotron J discharge $61484$, the reference cases in the present study. Among the LHD discharge $138675$ there is a co-ECCD injection that leads to the enhancement of the $n=1$ EPM, $n=2$ and $n=3$ TAEs. On the other hand, the ECCD is not injected in the Heliotron J discharge $61484$, showing unstable $n=1$ EPM and $n=2$ GAE. In addition to the reference cases, a set of equilibria are calculated modifying the rotational profile mimicking the effect of the ECCD for different current drive intensities and orientations \cite{48,7}. The electron density and temperature profiles are reconstructed by Thomson scattering data and electron cyclotron emission. Table~\ref{Table:1} and~\ref{Table:2} show the main parameters of the thermal plasma in Heliotron J and LHD reference cases, respectively.

\begin{table}[t]
\centering
\begin{tabular}{c c c c}
\hline
$T_{e0}$ (keV) & $n_{i0}$ ($10^{20}$ m$^{-3}$) & $\beta_{th,0}$ ($\%$) & B (T) \\ \hline
1 & 0.2 & 0.5 & 1.25 \\
\end{tabular}
\caption{Thermal plasma properties in the Heliotron J reference case (values at the magnetic axis). The first column is the thermal electron temperature, the second column is the thermal ion density, the third column is the thermal $\beta$ and the fourth column is the magnetic field intensity.} \label{Table:1}
\end{table}

\begin{table}[t]
\centering
\begin{tabular}{c c c c}
\hline
$T_{e0}$ (keV) & $n_{i0}$ ($10^{20}$ m$^{-3}$) & $\beta_{th,0}$ ($\%$) & B (T) \\ \hline
2.1 & 0.026 & 0.12 & 1.375 \\
\end{tabular}
\caption{Thermal plasma properties in the LHD reference case (values at the magnetic axis). The first column is the thermal electron temperature, the second column is the thermal ion density, the third column is the thermal $\beta$ and the fourth column is the magnetic field intensity.} \label{Table:2}
\end{table}

The nominal EP energy used in the simulations ($T_{f}$) is the energy resulting in an averaged Maxwellian energy equal to the average energy of a slowing-down distribution. For simplicity, no radial dependency of the EP energy is assumed in the Heliotron J simulations.

\begin{figure}[h!]
\centering
\includegraphics[width=0.45\textwidth]{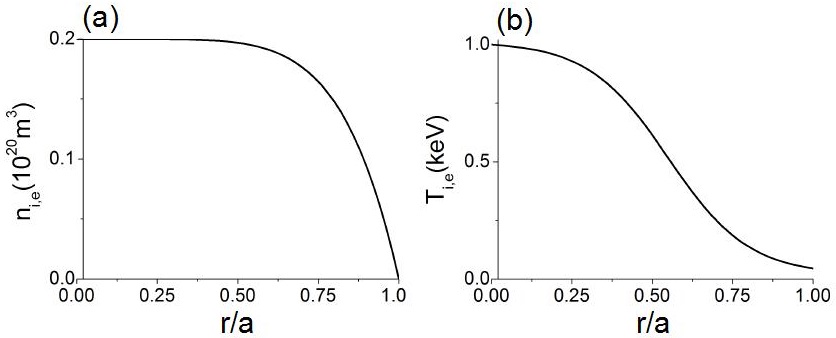}
\caption{Thermal plasma density (a) and thermal plasma temperature (b) in the Heliotron J model.}\label{FIG:1}
\end{figure}

\begin{figure}[h!]
\centering
\includegraphics[width=0.45\textwidth]{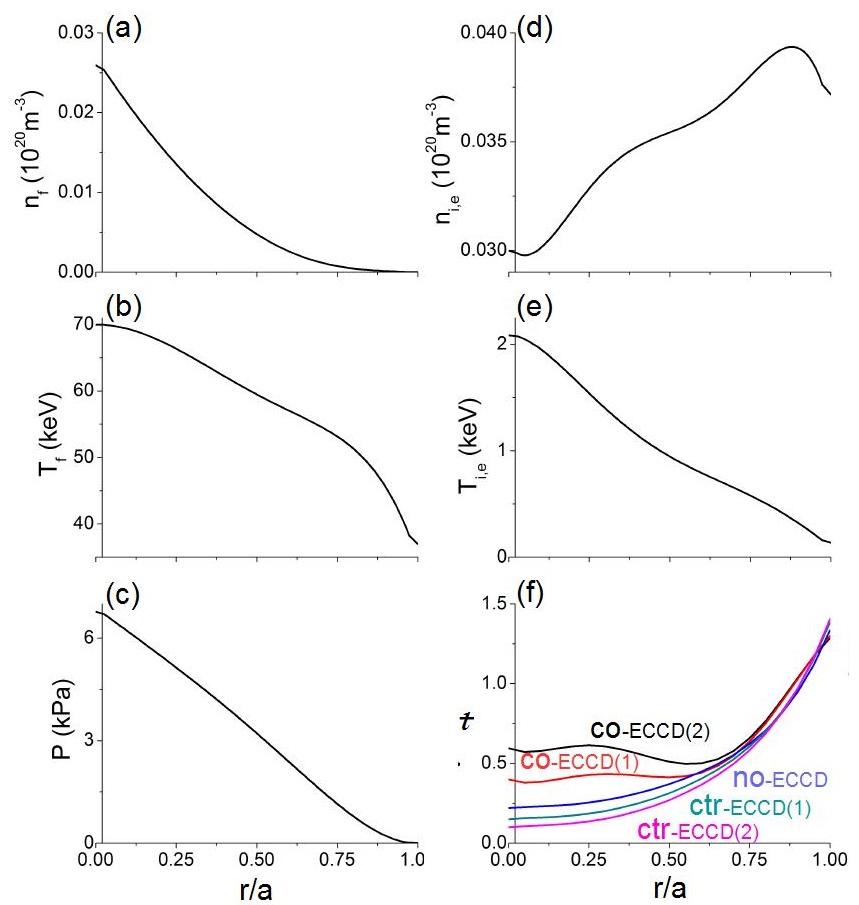}
\caption{EP density (a), EP temperature (b), total pressure (c), thermal plasma density (d), thermal plasma temperature (e) and rotational profile in the LHD cases analyzed (f).}\label{FIG:2}
\end{figure}

Figure~\ref{FIG:1} and ~\ref{FIG:2} show the thermal plasma and EP main profiles for the Heliotron J and LHD models, respectively. It should be noted that the effect of the equilibrium toroidal rotation is not included for simplicity.

\subsection{Simulations parameters}

The dynamic and equilibrium toroidal ($n$) and poloidal ($m$) modes in the simulations are shown in table~\ref{Table:3}. The simulations are performed with an uniform radial grid of 1000 points. In the following, the mode number notation is $n/m$, which is consistent with the $\rlap{-} \iota = n/m$ definition for the associated resonance. The number of equilibrium poloidal modes ($n=0$) required in the simulation relies upon the complexity of the device magnetic geometry, thus a larger number of modes are required to correctly reproduce the plasma stability as the device magnetic trap becomes more sophisticated. The selection of the equilibrium poloidal mode range is done in accordance with the VMEC code calculations. The range of dynamic poloidal modes for each toroidal family ($n=1$ to $3$) are chosen to include in the simulations all the resonant rational surfaces as well as the most important non resonant rational surfaces.

\begin{table}[t]
\centering
\begin{tabular}{c c c c}
\hline
$n$ & $m$ (Heliotron J) & $m$ (LHD) \\ \hline
0 & [0,14] & [0,10] \\
1 & [1,3] & [1,5] \\
2 & [2,6] & [2,10] \\
3 & ----- & [3,15] \\
\end{tabular}
\caption{Equilibrium ($n=0$) and dynamic ($n=1$ to $n=3$) toroidal and poloidal modes in the simulations for the Heliotron J and LHD models.} \label{Table:3}
\end{table}

The closure of the kinetic moment equations (equations (6) and (7) in the reference \cite{44}) breaks the MHD parities thus both parities must be included for all the dynamic variables. There is a detailed explanation of the kinetic closures in the references \cite{39,40,41}. The convention of the code with respect to the Fourier decomposition is, for example, the mode Fourier component of the mode $1/2 \to \cos(\theta + 2\zeta)$ and the mode $-1/-2 \to \sin(\theta + 2\zeta)$. The magnetic Lundquist number is assumed $S=5\cdot 10^6$.

\section{Stabilization of EPM/GAE in Heliotron J by ctr-ECCD injection \label{sec:HJ}}

\begin{figure}[h!]
\centering
\includegraphics[width=0.45\textwidth]{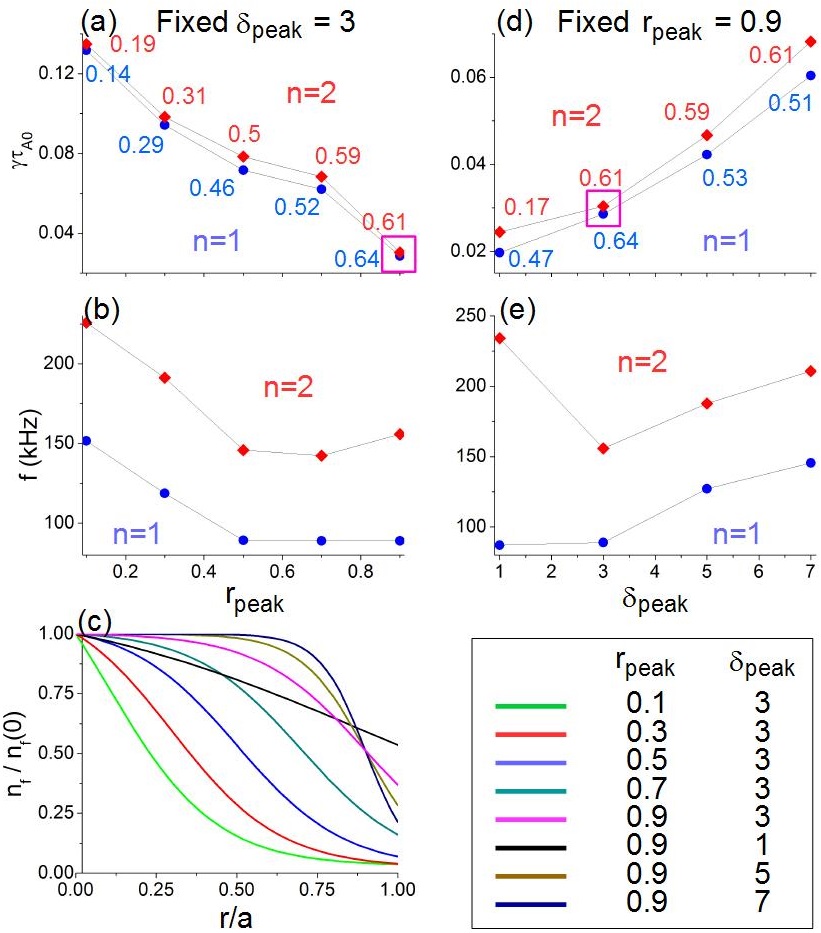}
\caption{Instability growth rate (a) and frequency (b) for different $r_{peak}$ values. EP particles profiles tested (c). Instabilities growth rate (d) and frequency (e) for different $\delta_{peak}$ values. The pink rectangles indicate the cases that better reproduce the experimental data. The labels in the graphs (a) and (d) show the radial location of the mode eigenfunction peak (normalized to the minor radius).}\label{FIG:3}
\end{figure}

In this section the stability of the Heliotron J plasma is analyzed with respect to the deformation of the rotational transform profile caused by the ECCD injection. Due to the lack of experimental data regarding the EP density and energy profiles, a preliminary study is required to identify the EP configuration that triggers instabilities consistent with the observations. That is to say, the simulations must reproduce the mode number, radial location and frequency range of the $n=1$ EPM and $n=2$ GAE measured in the experiment. The measured instabilities are triggered between the middle-outer plasma region, in the range of frequencies between $80$ to $100$ kHz for the $n/m = 1/2$ EPM and between $140$ to $160$ kHz for the $2/4$ GAE \cite{25}. First, the growth rate and frequency of the $n=1$ and $2$ instabilities are calculated using different EP density profiles (fixed $\beta_{f0} = 0.008$ and $T_{f0}=14$ keV, figure~\ref{FIG:3}). The EP density profile is given by the following analytical expression:
\begin{equation}
\label{EP_dens}
$$n_{f}/n_{f0}(r) = \frac{(0.5 (1+ \tanh(\delta_{peak} \cdot (r_{peak}-r))+0.02)}{(0.5 (1+\tanh(\delta_{peak} \cdot r_{peak}))+0.02)}$$
\end{equation}
The radial location of the profile gradient is controlled by the parameter ($r_{peak}$) and the flatness by ($\delta_{peak}$). Figure~\ref{FIG:3}c shows the EP density profiles tested. The simulations indicate a decrease of the instability growth rate as the EP density profile gradient is located closer to the plasma periphery (off-axis NBI injection, panels a and b) and the gradient is reduced (there is less free energy available to destabilize the EPM/GAE, panels d and e). The radial location of the mode eigenfuction peak (blue and red labels in panels a and d) moves outward as $r_{peak}$ increases, particularly if $\delta_{peak} = 3$. If $r_{peak} \ge 0.5$ there is a mode branch transition from an $n=1$ BAE and an $n=2$ TAE destabilized near the magnetic axis to an $n=1$ EPM and an $n=2$ GAE unstable in the middle-outer plasma region. The EP density profile that best reproduces the experimental observation is  highlighted by the pink rectangles, corresponding to an EP density profile with $r_{peak} = 0.9$ and $\delta_{peak} = 3$ (pink line in the fig~\ref{FIG:3}c). The simulations results show the $n=1$ EPM located at $r/a = 0.64$ and the $n=2$ GAE at $0.61$. It should be noted that the experimental observations indicate that the $2/4$ GAE is located around $r/a \approx 0.65$ although the $1/2$ EPM is located closer to the plasma periphery, at $r/a \approx 0.85$.

\begin{figure}[h!]
\centering
\includegraphics[width=0.45\textwidth]{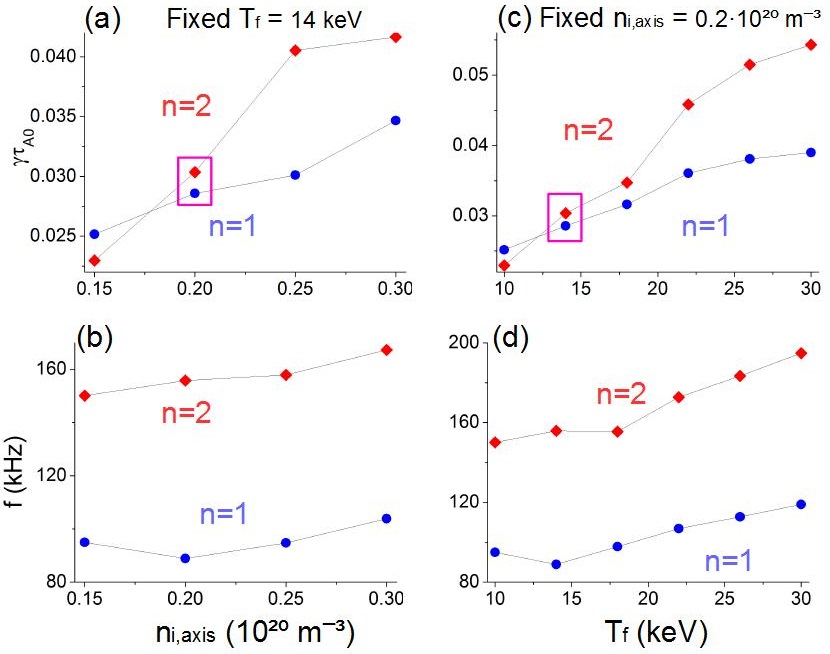}
\caption{Instabilities growth rate (a) and frequency (b) for different thermal ion densities (value at the magnetic axis). Instabilities growth rate (c) and frequency (d) for different EP energies. The pink rectangles indicate the cases that better reproduce the experimental data.}\label{FIG:4}
\end{figure}

Now, the resonance between the EP and the bulk plasma is analyzed with respect to the thermal ion density and the EP energy. The resonance is given by the ratio between the thermalized velocity of the EP ($\propto \sqrt{T_{f}}$) and the Alfven velocity ($\propto 1/\sqrt{n_{i}}$). Figure~\ref{FIG:4} shows the growth rate and frequency of the $n=1$ and $2$ modes calculated using different thermal ion densities (panels a and b) and EP energies (panels c and d) if the EP density profile is fixed ($r_{peak} = 0.9$ and $\delta_{peak} = 3$). If the thermal ion density or the EP energy increases (larger velocity ratio) the growth rate of the instability increases. The simulations with $n_{i,axis} = 0.2 \cdot 10^{20}$ m$^{-3}$ and $T_{f} = 14$ keV show the best agreement with respect to the frequency of the modes measured in the experiment, $89$ kHz for the $n=1$ mode and $150$ kHz for the $n=2$ mode (pink rectangle). The radial location of the modes eigenfunction peak is the same in all the simulations, $r/a = 0.64$ for the $n=1$ mode and $r/a=0.61$ for the $n=2$ mode. It should be noted that the measured averaged electron density in the discharge is around $1.0 \cdot 10^{19}$ m$^{-3}$ although in the simulations the assumed averaged plasma density is $1.7 \cdot 10^{19}$ m$^{-3}$. This value is obtained analyzing the effect of the thermal plasma density on the plasma stability, providing the best fit between Stellgap and FAR3d codes simulation results with respect to the Heliotron J measurement. The consequence is a slight enhancement of the resonance (see fig~\ref{FIG:4}a).

In summary, the model can reproduce the instabilities observed in the experiment although some discrepancies exist, caused by the lack of experimental data regarding the EP density and energy radial profiles, reducing the simulations accuracy.

\begin{figure}[h!]
\centering
\includegraphics[width=0.45\textwidth]{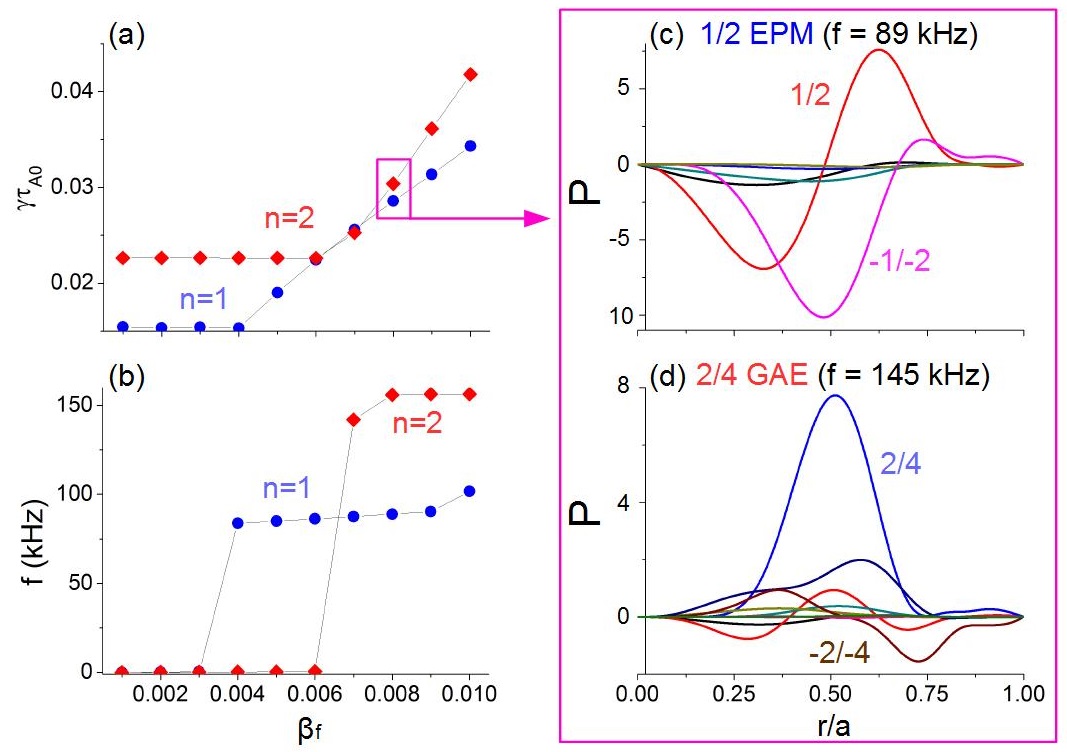}
\caption{Instability growth rate (a) and frequency (b) for different EP $\beta$ values. Eigen-function of the $n=1$ EPM (c) and $n=2$ GAE (d) if $\beta_{f} = 0.008$. The pink rectangles indicate the instability eigen-functions plotted.}\label{FIG:5}
\end{figure}

Next, the destabilization threshold of the $n=1$ EPM and $n=2$ GAE is analyzed with respect to the NBI injection intensity (EP $\beta$). Figure~\ref{FIG:5} shows the instability growth rate and frequency as the EP $\beta$ increases. In addition, the $n=1$ EPM and $n=2$ GAE eigen-function is plotted for an EP $\beta$ of $0.008$. The EP $\beta$ threshold to destabilize the $n=1$ EPM is $0.004$ and $0.007$ for the $n=2$ GAE (panels a and b). The dominant mode of the $n=1$ EPM is the $1/2$ and the $2/4$ for the $n=2$ GAE (panels c and d), consistent with the experimental data.

\begin{figure}[h!]
\centering
\includegraphics[width=0.45\textwidth]{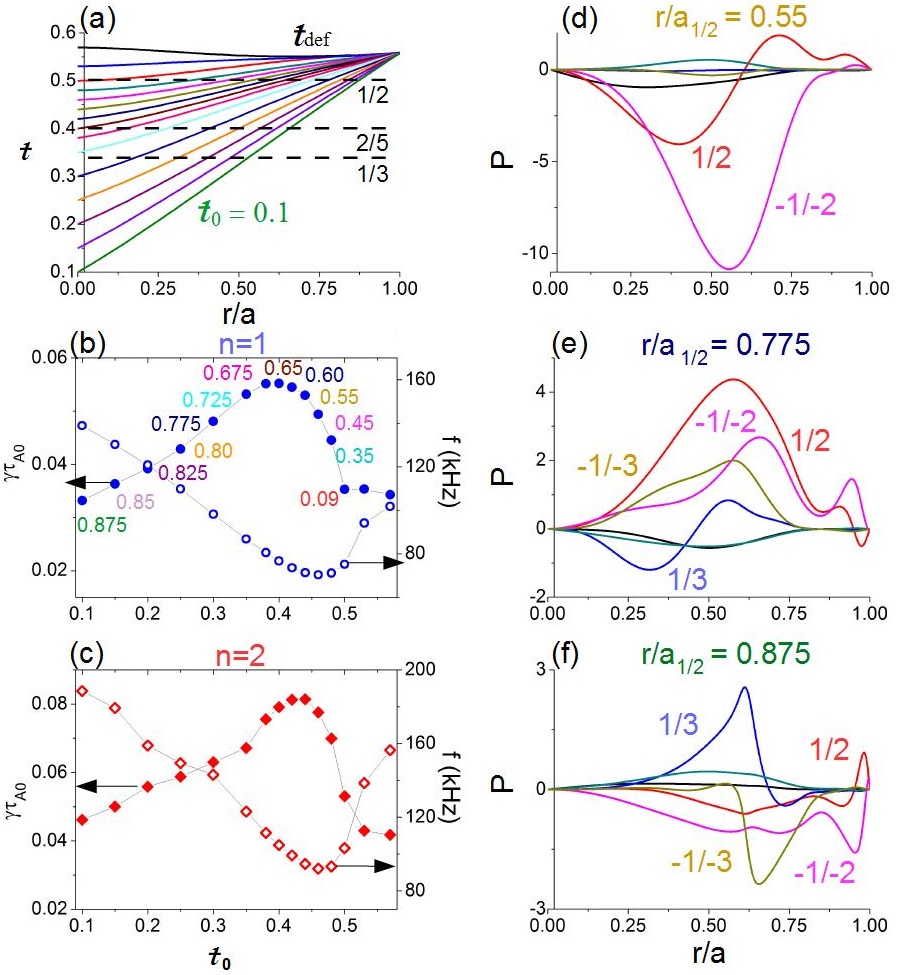}
\caption{(a) Rotational transform profile for different ctr-ECCD  injection intensities. Growth rate (solid circle/diamond) and frequency (open circle/diamond) of the $n=1$ EPM (b) and $n=2$ GAE (c) with respect to the rotational transform value at the magnetic axis. The dashed horizontal black lines in panel (a) indicate the radial location of the rational surfaces $1/2$, $2/3$ and $1/3$. The colored numbers (same color as the respective iota profile) in the panel (b) indicate the location of the $1/2$ rational surface (normalized minor radius). Eigen-function of the $n=1$ instability if the $1/2$ rational surface is located at (d) $r/a = 0.55$, (e) $0.775$ and (f) $0.875$.}\label{FIG:6}
\end{figure}

\begin{figure}[h!]
\centering
\includegraphics[width=0.45\textwidth]{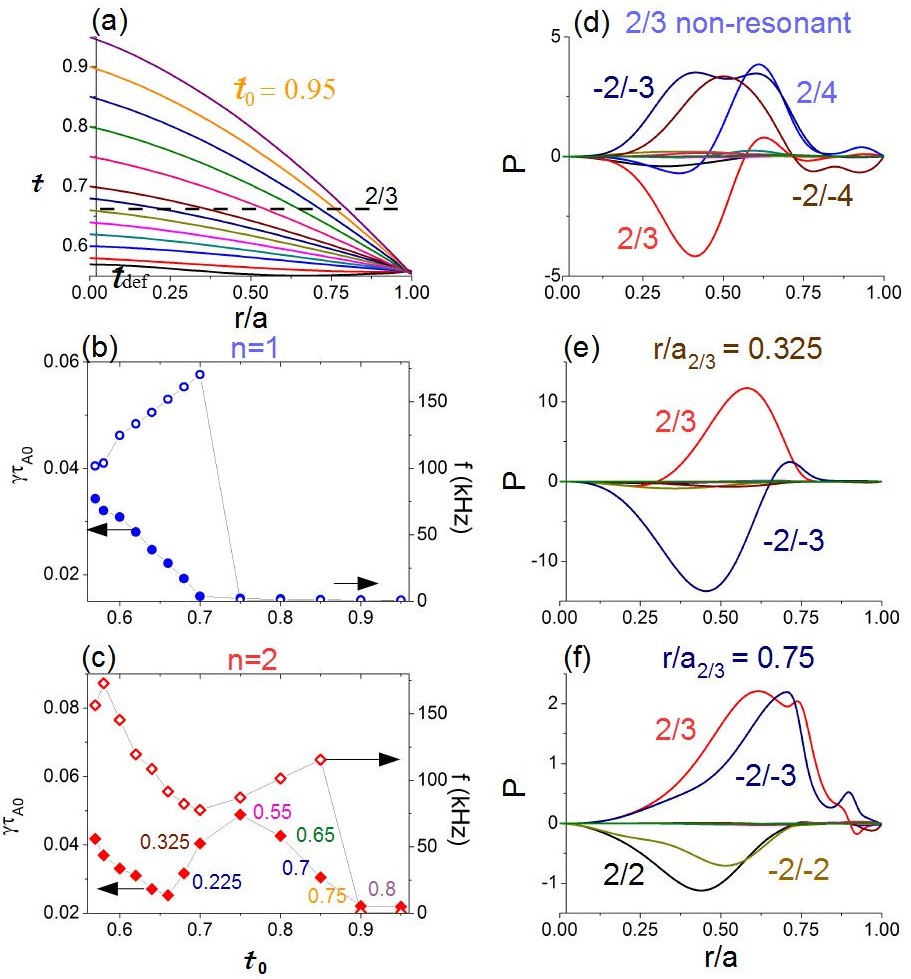}
\caption{(a) Rotational transform profile for different co-ECCD  injection intensities. Growth rate (solid circle/diamond) and frequency (open circle/diamond) of the $n=1$ EPM (b) and $n=2$ GAE (c) with respect to the rotational transform value at the magnetic axis. The dashed horizontal black line in panel (a) indicates the rational surface $2/3$. The colored numbers (same color as the respective iota profile) in the panel (b) indicate the location of the $2/3$ rational surface (normalized minor radius). Eigen-function of the $n=2$ instability if the $2/3$ rational surface is non resonant, (e) $r/a = 0.325$ and (f) $0.75$.}\label{FIG:7}
\end{figure}

The effect of the ECCD injection on the EPM/GAE stability is analyzed through the deformation induced in the rotational transform profile. Figures~\ref{FIG:6} and~\ref{FIG:7} indicate the deformation of the rotational transform profile for ctr- and co-ECCD injection intensities, the EPM/GAE growth rate and frequency as well as the modes eigenfunction.

The ctr-ECCD injection  generates a decrease of the rotational transform at the magnetic axis ($\rlap{-} \iota_{0}$, fig.~\ref{FIG:6}a), thus the magnetic shear between the inner and the outer plasma region increases. The growth rate of the $n=1$ EPM (fig.~\ref{FIG:6}b) and $n=2$ GAE (fig.~\ref{FIG:6}c) increases if $\rlap{-} \iota_{0}$ is between $[0.4 , 0.56]$, because the $1/2$ rational surface enters in the plasma and the mode destabilizing effect increases (colored number in the panel b), exceeding the stabilizing effect of the magnetic shear. Fig.~\ref{FIG:6}d shows the mode eigen-function if the $1/2$ rational surface is located at $r/a = 0.55$, slightly closer to the periphery with respect to the reference case. If $\rlap{-} \iota_{0} < 0.4$ the $1/2$ rational surface is located at $r/a > 0.65$ where the magnetic shear is stronger. Thus, the EPM/GAE growth rate decreases. Fig.~\ref{FIG:6}e and f show the mode eigen-function if $\rlap{-} \iota_{0}$ further decreases, so that the $1/2$ rational surface is located closer to the plasma periphery and the $1/3$  rational surface enters in the plasma, leading to a weaker destabilizing effect by the $1/2$. Consequently, the simulations reproduce the weakening of the EPM/GAE as the ctr-ECCD injection intensity increases (the modes growth rate is lower, consistent with the decrease of the modes amplitude measured in the experiment), as well as the threshold with respect to the ctr-ECCD intensity \cite{7}.

The co-ECCD injection produces an increase of the rotational transform at the magnetic axis and an enhancement of the magnetic shear (fig.~\ref{FIG:7}a). The growth rate of the $n=1$ EPM (fig.~\ref{FIG:7}b) decreases as $\rlap{-} \iota_{0}$ increases, leading to the mode stabilization if $\rlap{-} \iota_{0} \geq 0.75$, because the $1/2$ rational surface is non resonant and the energy channeled towards the $1/2$ mode is smaller. On the other hand, the growth rate of the $n=2$ GAE (fig.~\ref{FIG:7}c) decreases if $\rlap{-} \iota_{0} \le 0.66$, although it increases again between $\rlap{-} \iota_{0} = 0.68$ and $0.75$ because the rational surface $2/3$ enters in the plasma (see fig.~\ref{FIG:7}d and e). If $\rlap{-} \iota_{0}$ increase above $0.8$ the growth rate of the $n=2$ GAE decreases again because the $2/3$ textcolor{red}{rational surface is located at the plasma periphery where the magnetic shear is stronger} (see fig.~\ref{FIG:7}f), leading to the mode stabilization if $\rlap{-} \iota_{0} > 0.85$. In summary, the simulations reproduce the stabilizing effect of the co-ECCD injection observed in the experiment. In addition, the analysis also suggests the destabilizing effect of the $2/3$ rational surface, although there is no experimental evidence of the destabilization of a $2/3$ GAE in Heliotron J experiments.

\begin{figure}[h!]
\centering
\includegraphics[width=0.5\textwidth]{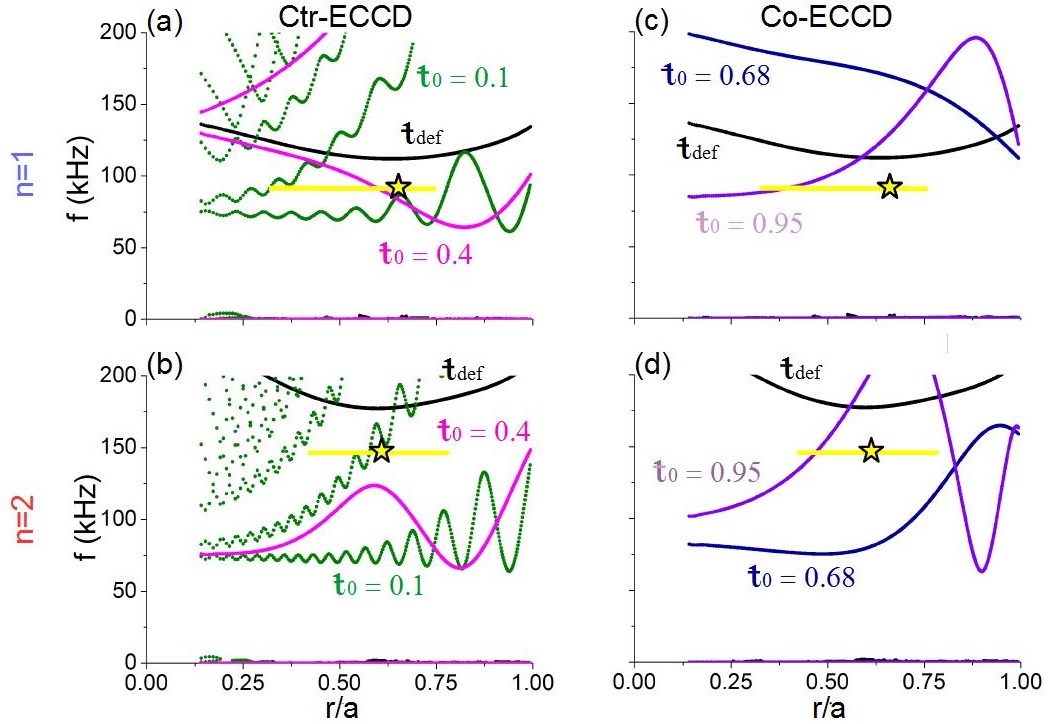}
\caption{Alfven gaps of the $n=1$ (a) and $n=2$ (b) modes in the ctr-ECCD equilibria if $\rlap{-} \iota_{0} = 0.4$ (purple line) and $0.1$ (green line). Alfven gaps of the $n=1$ (c) and $n=2$ (d) modes in the co-ECCD equilibria if $\rlap{-} \iota_{0} = 0.68$ (dark blue line) and $0.95$ (violet line). The reference case is indicated by the black line. The yellow stars indicate the radial location of the $1/2$ EPM and $2/4$ GAE and the solid yellow line the width of the mode eigen-function in the reference case.}\label{FIG:8}
\end{figure}

Figure~\ref{FIG:8} shows the Alfven gaps of the $n=1$ and $n=2$ toroidal families in the ctr-ECCD (fig.~\ref{FIG:8}a and b) and co-ECCD cases (fig.~\ref{FIG:8}c and d) for different $\rlap{-} \iota_{0}$. The case that shows the strongest enhancement of the continuum damping is the ctr-ECCD model with $\rlap{-} \iota_{0} = 0.1$, because the bandwidth of the Alfven gaps is smaller and there is a stronger interaction of the modes with the continuum (yellow stars show the radial location of the mode and the yellow lines the eigen-functions width). On the other hand, the co-ECCD cases show a weaker enhancement of the continuum damping as $\rlap{-} \iota_{0}$ increases with respect to the ctr-ECCD examples. 

In summary, the effect of the ctr-ECCD injection in Heliotron J leads to a large decrease of $\rlap{-} \iota_{0}$, thus the magnetic shear and the continuum damping increase at the middle-outer plasma region, leading to the stabilization of the $2/4$ GAE and a weaker $1/2$ EPM.

\section{Stabilization of EPM/TAE in LHD by ctr-ECCD injection \label{sec:LHD}}

\begin{figure*}[h!]
\centering
\includegraphics[width=0.8\textwidth]{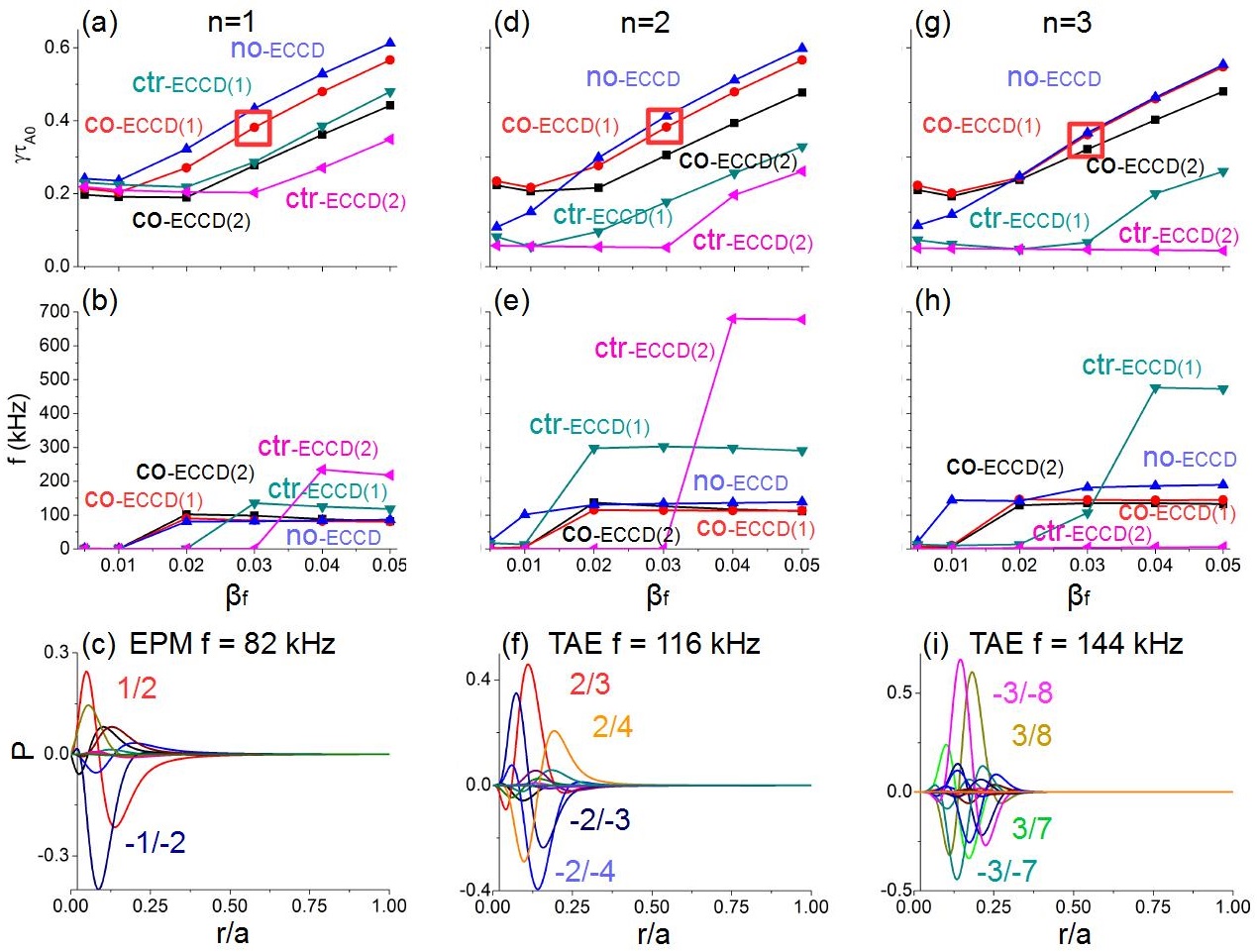}
\caption{Instability growth rate (a, d and g) and frequency (b, e and h) of the $n=1$ to $3$ instabilities for different EP $\beta$ values in the models with co-ECCD (black and red lines), no-ECCD (blue line) and ctr-ECCD (cyan and pink lines). Fig.~\ref{FIG:2}f indicates the $\rlap{-} \iota$ profile of the different ECCD scenarios. Eigen-function of the $n=1$ (c), $n=2$ (f) and $n=3$ (i) instabilities if $\beta_{f} = 0.03$. The red rectangles indicate the instability eigen-function plotted.}\label{FIG:9}
\end{figure*}

In this section the stability of the EPM/AEs is analyzed in LHD plasma where the ECCD is injected (the $\rlap{-} \iota$ profile of the different ECCD scenarios is shown in fig.~\ref{FIG:2}f). First, the $n=1$ EPM, $n=2$ and $n=3$ TAEs observed in the experiments are reproduced and the EP $\beta$ threshold is identified. Figure~\ref{FIG:9} shows the growth rate and frequency of the $n=1$ to $3$ modes if there is a co-ECCD, ctr-ECCD and no-ECCD injection for different EP $\beta$ values, as well as the modes eigen-function if $\beta_{f} = 0.03$. 

\begin{figure}[h!]
\centering
\includegraphics[width=0.3\textwidth]{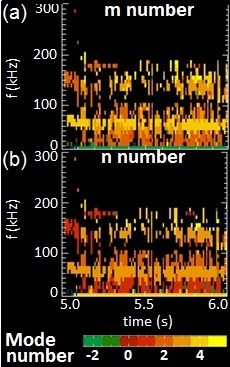}
\caption{(a) Poloidal and (b) toroidal mode number of the instabilities in the LHD co-ECCD discharge $138675$.}\label{FIG:10}
\end{figure}

The EP $\beta$ threshold required to destabilize the $n=1$ EPM and $n=2$ TAE is $0.04$, although the $n=3$ TAE is stable if EP $\beta < 0.05$ in the ctr-ECCD(2) case, higher with respect to the co-ECCD and no-ECCD cases where the EP $\beta$ threshold is $0.02$ (panels a, b, d, e, g and h). Also, above the EP $\beta$ destabilization threshold, the growth rate of the EPM/TAE is lower in the ctr-ECCD cases with respect to the co- and no-ECCD cases. Consequently, the EPM/AE stability improves in the ctr-ECCD cases. A $1/2$ EPM with a frequency of $82$ kHz (panel c), a $2/3-2/4$ TAE with a frequency of $116$ kHz (panel f) and a $3/7-3/8$ TAE with $f=144$ kHz are destabilized in the inner plasma. The AEs are triggered in the inner plasma because the density profile gradient of the EP density is located in the inner plasma. Figure~\ref{FIG:10} indicates the poloidal and toroidal mode number of the instabilities measured in the LHD discharge $138675$ using the Mirnov coils. In the frequency range below $60$ kHz, a robust $n=1$ and $m=2$ instability is observed in the experiment, pointing out that the frequency of the $1/2$ EPM reproduced in the simulations is overestimated around $20$ kHz. Also, in the frequency range of $80-110$ kHz a $n=2$ and $m=3-4$ instability is observed, comparable frequency with respect to the $2/3-2/4$ TAE in the simulations. A robust $n=3$ instability is also observed in the frequency range of $70$ kHz, although the simulations only reproduce the $n=3$ instability in the frequency range of $150$ kHz. It should be noted that the signal of the Mirnov coils is stronger for instabilities located closer to the plasma periphery with respect to the instabilities triggered closer to the plasma core, reason that could explain why the signal of the $3/4$ instability in the frequency range of $70$ kHz is stronger compared to the $n=3$ and $m>4$ instability in the frequency range of $150$ kHz, that is to say, the $3/4$ instability is located closer to the Minor coils. In addition, the simulations for the $n=3$ toroidal family indicate the destabilization of a $3/7-3/8$ TAE, although the measurements shows an instability with an $m=5$. Again, the discrepancy between experimental data and simulations could be caused by the EP density/energy and  iota profiles assumed in the simulations. Nevertheless, all the modes analyzed indicate the same trend, an improvement of the AE/EPM stability in the ctr-ECCD cases.

\begin{figure}[h!]
\centering
\includegraphics[width=0.5\textwidth]{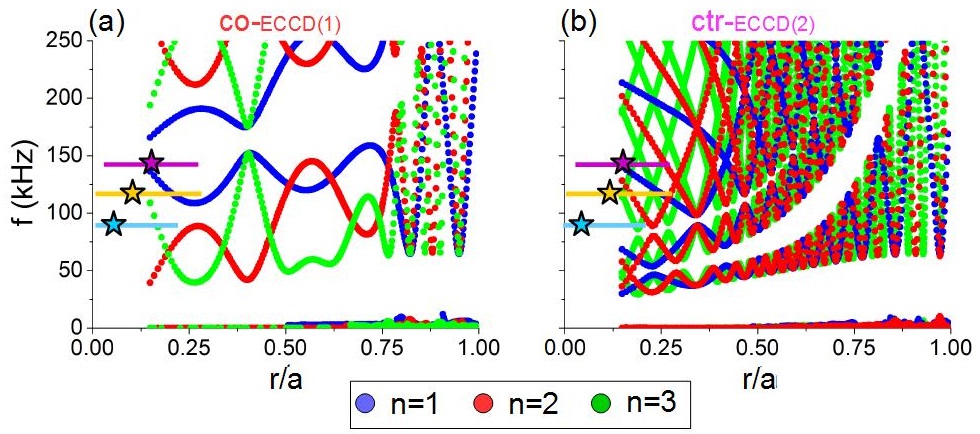}
\caption{Alfven gaps in the models co-ECCD(1) (a) and ctr-ECCD(2) (b) for the $n=1$ (blue dots), $n=2$ (red dots) and $n=3$ (green dots) modes. Fig.~\ref{FIG:2}f indicates the $\rlap{-} \iota$ profile of the different ECCD scenarios. The cyan star indicate the eigen-function local maxima of the $1/2$ EPM (fig 9c), the orange star the $2/3-2/4$ TAE (fig 9f) and the purple star the $3/7-3/8$ TAE (fig 9i) in the Alfven gaps. The solid color lines indicate the width of the modes eigen-function. }\label{FIG:11}
\end{figure}

Figure~\ref{FIG:11} shows the Alfven gaps of the cases with co-ECCD and ctr-ECCD injection. The model ctr-ECCD(2) (panel b) shows an enhancement of the continuum damping with respect to the co-ECCD(1) case (panel a), particularly in the inner-middle plasma region and the frequency ranges where the $1/2$ EPM, $2/3-2/4$ TAE and $3/7-3/8$ TAE are unstable in the co-ECCD and the no-ECCD cases.

\begin{figure}[h!]
\centering
\includegraphics[width=0.5\textwidth]{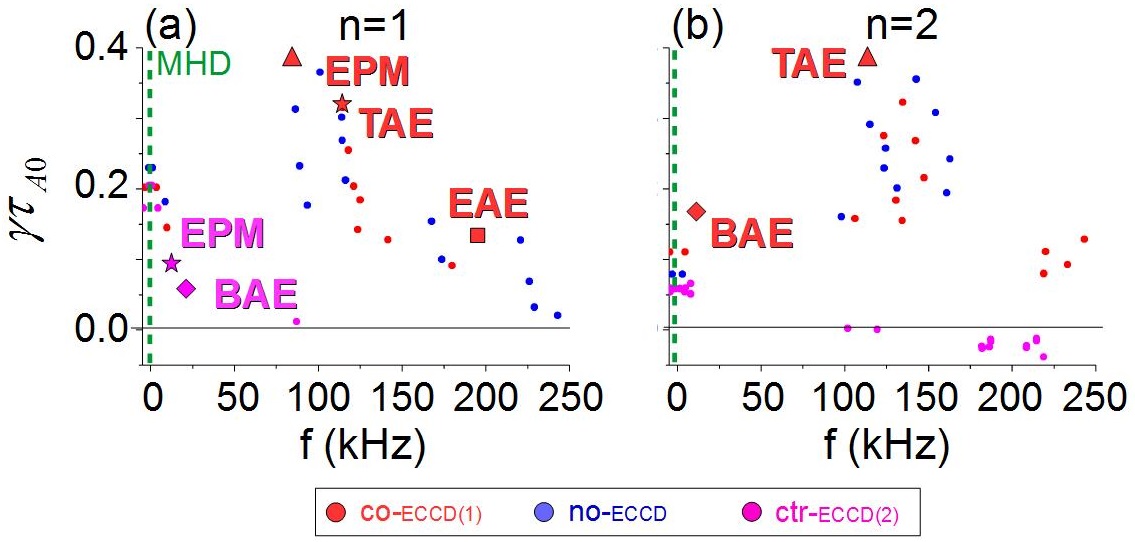}
\caption{Sub-dominant modes of the $n=1$ (a) and $n=2$ (b) toroidal families for the cases co-ECCD(1) (red dots), no-ECCD (blue dots) and ctr-ECCD(2) (pink dots). The dashed green lines indicate the range of frequency of the pressure gradient driven modes. The solid line indicates the threshold between stable modes (negative growth rate) and unstable modes (positive growth rate). The red triangles show the modes with the largest growth rate. The red star, diamond and square indicate the most unstable sub-dominant modes  in the co-ECCD(1) case. The pink star and diamond show the most unstable sub-dominant modes in the ctr-ECCD(2) case.}\label{FIG:12}
\end{figure}

To confirm the improvement of the EPM/AE stability in the cases with ctr-ECCD injection, figure~\ref{FIG:12} shows the sub-dominant modes calculated for the models co-ECCD(1), no-ECCD and ctr-ECCD(2) if the EP $\beta = 0.03$. The data of the analysis for the $n=3$ modes is not included because no sub-dominant modes  are identified. The $1/2$ EPM and the $2/3-2/4$ TAE (red triangles) are unstable in the co-ECCD(1) and no-ECCD cases although stable in the ctr-ECCD(2) case. In addition, several sub-dominant modes are destabilized in the co-ECCD(1) and no-ECCD cases, as a $n=2$ BAE with $f=13$ kHz (red diamond), a $n=1$ TAE with $f = 115$ kHz (red star) and a $n=1$ EAE with $f = 197$ kHz (red square). Regarding the ctr-ECCD(2) case, two sub-dominant modes are also unstable, a $n=1$ EPM with $f=5$ kHz (pink star) and a $n=1$ BAE with $f=20$ kHz (pink diamond).

\begin{figure}[h!]
\centering
\includegraphics[width=0.5\textwidth]{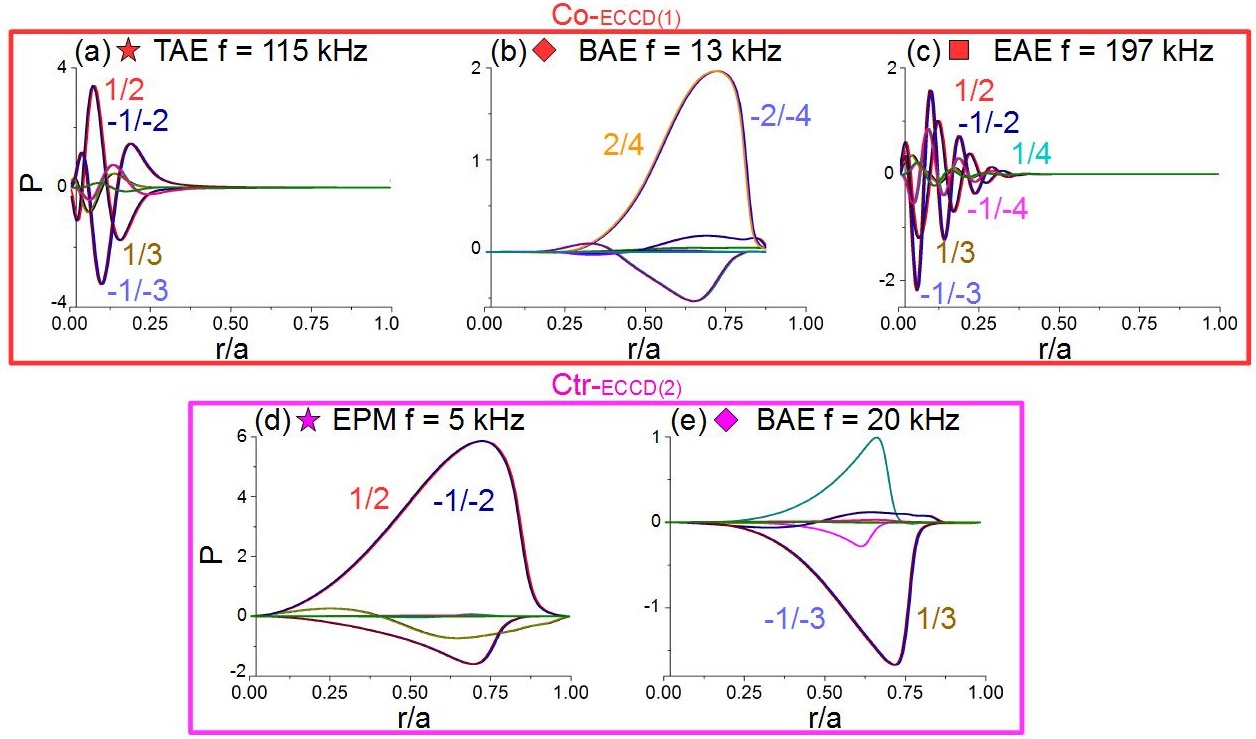}
\caption{Eigen-function of the $n=1$ TAE (a), $n=2$ BAE (b) and $n=1$ EAE (c) in the co-ECCD(1) model. Eigen-function of the $n=1$ EPM (d) and $n=1$ BAE (e) in the ctr-ECCD(2) model.}\label{FIG:13}
\end{figure}

Figure~\ref{FIG:13} shows the eigen-function of the most unstable sub-dominant modes in the co-ECCD(1) and ctr-ECCD(2) cases. The unstable modes in the co-ECCD(1) case are a $1/2 - 1/3$ TAE destabilized in the inner plasma (panel a), a $2/4$ BAE in the middle-outer plasma (panel b) and a $1/2-1/4$ EAE in the inner plasma region (panel c). If the results of the sub-dominant mode analysis are compared with the experimental data (fig~\ref{FIG:10}), there is an instability with $n=1$ and $m=2$ with $f \approx 110-130$ kHz as well as an $n=1$ and a $m=2-4$ with $f \approx 180-200$ kHz. On the other hand, no $n=2$ instability is observed in the frequency range of the $10-20$ kHz. The instabilities in the ctr-ECCD(2) case are a $1/2$ EPM (panel d) and a $1/3$ BAE (panel e) in the middle-outer plasma region, also consistent with the frequency range of the instabilities observed in the experiment (data not shown).

In summary, the co-ECCD injection in LHD discharges can trigger the AE/EPM because the stabilizing effect of the continuum damping is weakened. On the other hand, the ctr-ECCD injection leads to a stronger continuum damping and the stabilization of the $n=1$ EPM, $n=2$ TAE and $n=3$ TAE.

\section{Conclusions and discussion \label{sec:conclusions}}

A set of linear simulations are performed by the FAR3d code studying the effect of the ECCD injection on the Heliotron J and LHD plasma stability. The simulation results are compared with the experimental data showing a reasonable agreement with respect to the instability mode number, radial location and frequency range.

The stability of the EPM/AE of Heliotron J plasma is analyzed for different configurations where the $\rlap{-} \iota$ profile is modified, increasing/decreasing $\rlap{-} \iota_{0}$ mimicking the effect of the ECCD injection. The simulations show an improvement of the EPM/AE stability if the magnetic shear enhances as $\rlap{-} \iota_{0}$ increases (co-ECCD injection), although only below a given threshold if $\rlap{-} \iota_{0}$ decreases (ctr-ECCD injection). The $\rlap{-} \iota_{0}$ threshold in the simulations with ctr-ECCD injection is closely linked to the $1/2$ rational surface, entering in the plasma and exceeding the stabilizing effect of the magnetic shear. A further decrease of the $\rlap{-} \iota_{0}$ leads to a $1/2$ rational surface located at the plasma periphery where the magnetic shear is strong enough to stabilize the EPM/AE. It should be noted that a similar threshold is observed in Heliotron J discharges with ctr-ECCD injection, because the amplitude of the EPM/AE decreases only if the ECCD is above a given current intensity. On the other hand, no threshold is observed in discharges with co-ECCD injection with respect to the current intensity. In addition, the application of ECCD also leads to an enhancement of the continuum damping.

The simulation results of LHD discharges with ECCD injection indicate the further destabilization of EPM/AEs in the cases with co-ECCD injection and the stabilization in the cases with ctr-ECCD injection. The EPM and TAE observed in the experiment are destabilized in the inner plasma region where the continuum damping is enhanced as the $\rlap{-} \iota_{0}$ decreases due to the ctr-ECCD injection. It should be noted that the magnetic shear in the inner plasma is weakly affected by the ECCD injection, thus the main stabilizing effect on the EPM/AE is caused by the continuum damping. Consequently, the EP $\beta$ threshold to destabilize EPM/AEs is higher in LHD discharges with ctr-ECCD injection with respect to shots without ECCD or co-ECCD injection, thus the plasma stability improves. 

The ECCD is a useful tool in Heliotron J and LHD discharges to modify the rotational transform and improve the EPM/AE stability. In both discharges the ctr-ECCD injection causes a decrease of $\rlap{-} \iota_{0}$, enhancing the stabilizing effect of the continuum damping and the magnetic shear. Nevertheless, the dominant stabilizing effect depends on the radial location and the frequency range of the unstable modes. The continuum damping is more efficient to stabilize EPM/AEs located in the inner-middle plasma region with frequencies above the BAE frequency range. On the other hand, if the unstable modes are located between the middle-outer plasma region, the magnetic shear is the main stabilizing factor.

The injection of ECCD in nuclear fusion devices, particularly in Stellarators, combined with other sources of non inductive currents such as the neutral beam current drive (NBCD) or the low hybrid (LH), can lead to an improved operation scenario with respect to the EPM/AE stability, not accessible using the standard magnetic field configuration generated by the coils. Dedicated experiments in Heliotron J and LHD will be performed in future campaigns to analyze in more detail the optimization trends linked to non inductive currents.

\ack

The authors want to thank the LHD technical staff for their contributions in the operation and maintenance of LHD. This work was supported in part by JSPS Core-to-Core Program, A. Advanced Researcher Network, by NIFS under contracts NIFS07KLPH004 and NIFS17KLPR038 and by the Comunidad de Madrid under the project 2019-T1/AMB-13648. The work of an author (K. Nagasaki) was supported by the NIFS Collaborative Research Program (NIFS08KAOR010, NFIS10KUHL030), a Grant-in-Aid for Scientific Research (B),KAKENHI, and "PLADyS", JSPS Core-to-Core Program, A. Advanced Research Networks. The authors also want to acknowledge Y. Suzuki for fruitful discussion.

\hfill \break

\end{document}